\documentclass[useAMS,usenatbib]{mn2e}

\usepackage[psamsfonts]{amssymb}
\usepackage[dvips]{graphicx}
\usepackage{amsmath,alltt}
\usepackage{multirow}
\usepackage{rotating}
\usepackage{lscape}

\title[Quantifying the Universality of the Stellar IMF]{Quantifying
  the Universality of the Stellar Initial Mass Function in Old 
  Star Clusters}
\author[Leigh et al.]{Nathan
  Leigh$^{1,2}$, Stefan Umbreit$^{3,4}$, Alison Sills$^{1}$, Christian
  Knigge$^{5}$, 
\newauthor
  Guido de Marchi$^{2}$, Evert Glebbeek$^{1,6}$, Ata Sarajedini$^{7}$\thanks{E-mail: 
    leighn@mcmaster.ca (NL); s-umbreit@northwestern.edu (SU); 
    asills@mcmaster.ca (AS1); christian@astro.soton.ac.uk (CK);
    e.glebbeek@astro.ru.nl (EG); gdemarchi@esa.rssd.int (GD);
    ata@astro.ufl.edu (AS2)}\\
$^{1}$Department of Physics and Astronomy, McMaster University,
1280 Main St. W., Hamilton, ON, L8S 4M1, Canada \\
$^{2}$European Space Agency, Space Science Department, Keplerlaan 1,
2200 AG Noordwijk, The Netherlands \\
$^{3}$Center for Interdisciplinary Exploration and Research in 
Astrophysics (CIERA), Northwestern University, Evanston, IL 60208, USA \\
$^{4}$Department of Physics and Astronomy, 
Northwestern University, Evanston, IL 60208, USA \\
$^{5}$School of Physics and Astronomy, University of Southampton,
Highfield, Southampton, SO17 1BJ, United Kingdom \\
$^{6}$Department of Astrophysics/IMAPP, Radboud University Nijmegen, 
P.O. Box 9010, 6500 GL Nijmegen, The Netherlands \\
$^{7}$Department of Astronomy, University of Florida, Gainesville, FL
32611, USA}
\begin{document}

\pagerange{\pageref{firstpage}--\pageref{lastpage}} \pubyear{2010}

\maketitle

\label{firstpage}

\begin{abstract}
We present a new technique to quantify
cluster-to-cluster variations in the observed present-day stellar mass
functions of a large sample of star clusters.  Our method quantifies these 
differences as a function of both the stellar mass and the total cluster 
mass, and offers the advantage that it is insensitive to the precise 
functional form of the mass function.  
We applied our technique to data taken from the Advanced Camera for 
Surveys Survey for Globular 
Clusters, from which we obtained completeness-corrected stellar mass
functions in the mass range 0.25-0.75 M$_{\odot}$ for a sample of 27
clusters.  The results of our observational analysis were then compared 
to Monte Carlo simulations for globular cluster evolution spanning a range of
initial mass functions, total numbers of stars, concentrations, and 
virial radii.  

We show that 
the present-day mass functions of the clusters in our sample can 
be reproduced by assuming an universal 
initial mass function for all clusters, and that the cluster-to-cluster 
differences are consistent with what is expected from 
two-body relaxation.  A more complete exploration of the initial 
cluster conditions will be needed in future studies to better constrain 
the precise functional form of the initial mass function.  This study 
is a first step toward using our technique to constrain the dynamical 
histories of a large sample of old Galactic star clusters and, by extension, 
star formation in the early Universe.  
\end{abstract}


\section{Introduction} \label{intro}
It is now thought that most, if not all, of the stars in our Galaxy were
born in star clusters \citep[e.g.][]{lada95, lada03, mckee07}.  And yet, 
there remain several key details of the star formation process that are still
not understood.  Part of the 
problem lies in the fact that populations of young stars are typically
hidden by a dense veil of optically-thick gas and dust.  This
prevents the escape of most of the light produced by infant stars, and
often renders these regions difficult to observe \citep[e.g.][]{grenier05,
  lada07}.  Most of 
these clusters are sparsely populated and are of relatively low mass (M
$\lesssim$ 10$^4$ M$_{\odot}$) \citep[e.g.][]{lada85}.  They are also
very young since clusters of such low mass are unlikely to survive for more than
1 Gyr \citep[e.g.][]{portegieszwart10}.    

At the other end of the cluster mass spectrum, most massive star clusters (M
$\gtrsim$ 10$^4$ M$_{\odot}$) in our Galaxy tend to be at least a few
Gyrs old, and 
in many cases are nearly as old as the Universe itself
\citep[e.g.][]{harris96, deangeli05}.  These clusters have the advantage that 
they are no longer obscured by the primordial gas 
from which they formed, however they are a
dynamically active environment.  As a result, the conditions present at the
time of their formation have now been largely erased
\citep[e.g.][]{portegieszwart01, hurley05, murray09}.  This presents a 
considerable challenge for studying star formation in the regime of
cluster masses and metallicities that characterize Milky Way globular 
clusters.  This is unfortunate since these old star clusters contain 
the fossil record of a very 
early episode of star formation in the Universe, and are the only
means of studying it locally in massive star clusters.

One of the primary
observational tests for star formation theories is the stellar initial 
mass function (IMF).  Current observational evidence suggests that the
IMF is very similar in different regions of our Galaxy,
including the disk and young star clusters
\citep[e.g.][]{elmegreen99, kroupa11}.  However, this is still being debated
throughout the literature \citep[e.g.][]{scalo98, parravano11}.  
Different star formation theories tend to predict different IMFs.  These
vary with the properties of the gas clouds from which the
stars are born, including density, temperature and composition
\citep[e.g.][]{elmegreen01, bonnell07, mckee07, kroupa11}.  

Given the sensitive nature of the observations, a
large sample of IMFs spanning the entire range of cluster properties
exhibited by star clusters in the Milky Way, 
including total mass and chemical composition, has yet to be
compiled.  
This is a sorely needed step in order to advance our understanding
of star formation by providing direct comparisons between observations
and theoretical 
predictions.  This is especially true of massive, metal-poor star 
clusters since we are particularly 
lacking observations of IMFs in this regime of cluster masses and
metallicites \citep[e.g.][]{mckee07,
  portegieszwart10}.   Important steps in this direction were 
recently taken by \citet{demarchi10} and \citet{paust10}, who studied
the present-day mass 
functions of a large sample of Galactic clusters
and considered the effects of the cluster dynamics in modifying them
from their primordial forms. 

For the very first time, the Advanced Camera for Surveys (ACS) Survey for Globular Clusters has
provided photometry for a large sample of Milky Way globular clusters 
(GCs) that reaches down to unprecedented faint magnitudes.  This offers a 
large sample of current stellar mass functions spanning the stellar mass range
$\approx 0.2 - 0.8$ M$_{\odot}$.  All of the
clusters are massive and very old, with total masses and ages
ranging from $\approx$ 10$^4$ - 10$^6$ M$_{\odot}$ and $\approx$ 10-12 Gyrs,
respectively 
\citep{harris96, deangeli05}.  This has allowed significant time for
their stellar mass functions to have been modified from their
primordial forms due to both stellar evolution and stellar dynamics.
However, most of the processes responsible for this evolution are now
largely understood.  Therefore, in principle, it is possible to use
current observations of old star clusters together with theoretical models
for their evolution to extrapolate backwards in time and indirectly
probe their IMFs.

For most of the life of a massive star cluster, two-body relaxation is the 
dominant physical mechanism driving its evolution
\citep[e.g.][]{henon60, spitzer87, heggie03, gieles11}.  
The term describes the cumulative effects of long-range gravitational
interactions that occur between pairs of
stars, which act to alter their orbits within the cluster.  This results in a
phenomenon known as mass segregation, which is the tendency for
heavier stars to accumulate in the central cluster regions and
low-mass stars to be dispersed to wider orbits.  This mechanism also causes 
stars to escape from their host cluster, with 
the probability of ejection increasing with decreasing cluster mass.  
Therefore, two-body relaxation acts to slowly modify the distribution of stellar 
masses within clusters, and can cause very dynamically evolved clusters to 
appear severely depleted of their low-mass stars.  
Evidence in favour of this process having actually
occurred in real star clusters has been reported by several authors
\citep[e.g.][]{vonhippel98, demarchi10}.

A number of theoretical studies have been conducted to learn how 
the evolution of the stellar mass function (MF) in GCs is affected 
by two-body relaxation, stellar evolution, 
disc shocking, and tidal effects from the Galaxy (see \citet{baumgardt03} 
for a detailed review).  In the absence of these effects, we expect the MF 
to continually rise toward lower stellar masses.  
By performing a series of N-body simulations, 
\citet{vesperini97} showed that tidal effects from the Galaxy, disc shocking, 
and a higher initial central concentration all act to increase the rate of 
stellar evaporation, and 
accelerate the depletion of preferentially low-mass stars.  These results 
were confirmed and built upon by \citet{baumgardt03} who showed that the 
depletion of low-mass stars can be sufficiently dramatic to change the 
sign of the slope of the MF at the low-mass end.  
Interestingly, these results were not 
supported by the observational study of \citet{demarchi07}.  These authors 
analyzed the MFs in a sample of 20 Galactic GCs, and found that the slope of the 
MF decreases with increasing central concentration.  They argued that 
this contradicts what is expected from theory since two-body relaxation is 
responsible both for increasing the central density and flattening the MF at 
the low-mass end.  In an effort to explain this, they suggested that 
many of the clusters in their sample could be post-core collapse, and therefore 
had much higher central densities in the past.  Alternatively, 
\citet{marks08} argued that this can be explained by residual gas-expulsion 
from initially mass segregated clusters \citep[e.g.][]{tutukov78}, and cautioned 
that unresolved binaries could also be contributing.

Several theoretical studies have also been conducted to study the 
dynamical histories of individual globular clusters \citep[e.g.][]{heggie08, 
heggie09}.  For example, \citet{zonoozi11} recently performed the first ever 
direct N-body simulations of a Milky Way (MW) GC over its entire lifetime.  This 
was done for the distant GC Palomar 14, which has an unusually low-density and 
large radius.  The emphasis of this paper is to use the ensemble information 
of many GCs to learn about the universality of the IMF in old massive star clusters.  
Individual cases, in particular Pal 14, are often chosen for their peculiar 
characteristics and may not be representative of the bulk of the GCs in the MW.

In this paper, we present a new technique to quantify
cluster-to-cluster variations 
in the observed stellar mass functions of a large sample of clusters
spanning a diverse range of properties.  Our method offers the 
advantage that it is insensitive to the precise functional form of 
the MF.  We have applied it to a sample of 27 MFs 
taken from the ACS Survey for Globular Clusters \citep{sarajedini07}.  
Can the present-day MFs be explained by an universal IMF and stellar evaporation 
induced by two-body relaxation?  Or are cluster-specific IMFs needed 
to reproduce the observed MFs?  To address these questions, 
we compared the results of our observational analysis to 
268 Monte Carlo simulations for GC evolution.  
The models spanned a range of initial masses, virial radii, central
concentrations and IMFs.  Therefore, by evolving all of these models to 
the current ages of the GCs in our sample and comparing the resulting 
MFs to the present-day observed ones, we have quantified the 
dynamical evolution of the MFs in our observed sample.  This has 
allowed us to take the first steps toward constraining both the exact
functional forms of the IMFs of MW GCs and 
the conditions present at the time of their formation.

In Section~\ref{method}, we present our sample of observed stellar mass
functions and describe both our technique for analyzing the
observations and the models for globular cluster evolution to which
they are compared.  The results of our analysis of
the ACS observations are presented in Section~\ref{results}, along
with an example comparison between the observations and the models.  This 
example demonstrates how our method can be used to compare a large 
number of observed MFs to analogous samples of simulated MFs.  
Finally, we discuss in Section~\ref{discussion} the implications of
our results for the conditions present in our observed clusters at the 
time of their formation and the role played by two-body relaxation in 
modifying the 
stellar MF to its present-day 
form.  

\section{Method} \label{method}

In this section, we describe how we acquired our sample of mass
functions from the ACS data, as well as the Monte Carlo simulations
for globular cluster evolution used for comparison to the observations.

\subsection{The Data} \label{data}

The data used in this study was taken from the sample of 35 MW GCs 
used in \citet{leigh11}, which was in turn taken 
from the ACS Survey for Globular Clusters
\citep{sarajedini07}.\footnote[1]{The
data can be found at http://www.astro.ufl.edu/$\sim$ata/public\_hstgc/.}
The ACS Survey provides unprecedented deep photometry in the F606W ($\approx$
V) and F814W ($\approx$ I) filters
that is nearly complete down to $\approx 0.2$ M$_{\odot}$.  In other
words, the colour-magnitude diagrams (CMDs) extend reliably from the
horizontal branch all the way down to about 7 magnitudes below the main-sequence
turn-off (MSTO).  A list of the GCs used in this study is shown in 
Table~\ref{table:list} along with their core radii (r$_c$), half-mass 
radii (r$_h$), central luminosity densities ($\rho_0$), and central 
concentration parameters (c).  
These were taken directly from \citet{harris96}, with the exception of 
the core and half-mass radii.  The latter quantities are 
given in parsecs and were calculated using the distance modulii and 
extinction corrections provided in \citet{harris96}.  

Each cluster was centred in the ACS field, which
extends out to several core radii from the cluster
centre in most cases.  Coordinates for the cluster centres were
taken from 
\citet{goldsbury10}.  These authors found their centres by fitting
a series of ellipses to the density distributions within the inner 2'
of the cluster centre, and computing an average value.  


\begin{table}
\centering
\caption{List of globular clusters and their structural parameters
  \label{table:list}}
\begin{tabular}{|c|c|c|c|c|c|}
\hline
Cluster  &  Alternate  &    r$_c$    &    r$_h$    &           $\rho_0$           &  c   \\
  ID     &    ID       &   (in pc)   &   (in pc)   &  (in L$_{\odot}$ pc$^{-3}$)  &      \\   
\hline
 104 &    47 Tuc    &  0.47  &  4.11  & 4.88 & 2.07 \\
1261 &              &  1.66  &  3.22  & 2.99 & 1.16 \\
2298 &              &  0.97  &  3.08  & 2.90 & 1.38 \\
4147 &              &  0.51  &  2.70  & 3.63 & 1.83 \\
4590 &      M 68    &  1.73  &  4.51  & 2.57 & 1.41 \\
5024 &      M 53    &  1.82  &  6.80  & 3.07 & 1.72 \\
5272 &       M 3    &  1.10  &  6.84  & 3.57 & 1.89 \\
5286 &              &  0.95  &  2.48  & 4.10 & 1.41 \\
5904 &       M 5    &  0.96  &  3.85  & 3.88 & 1.73 \\
5927 &              &  0.94  &  2.46  & 4.09 & 1.60 \\
5986 &              &  1.43  &  2.97  & 3.41 & 1.23 \\
6093 &      M 80    &  0.44  &  1.78  & 4.79 & 1.68 \\
6121 &      M 4     &  0.59  &  2.18  & 3.64 & 1.65 \\
6171 &     M 107    &  1.04  &  3.21  & 3.08 & 1.53 \\
6205 &      M 13    &  1.29  &  3.51  & 3.55 & 1.53 \\
6218 &      M 12    &  1.11  &  2.49  & 3.23 & 1.34 \\
6254 &      M 10    &  0.98  &  2.49  & 3.54 & 1.38 \\
6304 &              &  0.36  &  2.43  & 4.49 & 1.80 \\
6341 &      M 92    &  0.63  &  2.45  & 4.30 & 1.68 \\
6535 &              &  0.71  &  1.68  & 2.34 & 1.33 \\
6584 &              &  1.02  &  2.86  & 3.33 & 1.47 \\
6637 &      M 69    &  0.84  &  2.15  & 3.84 & 1.38 \\
6779 &      M 56    &  1.21  &  3.02  & 3.28 & 1.38 \\
6838 &      M 71    &  0.74  &  1.96  & 2.83 & 1.15 \\
6934 &              &  1.00  &  3.14  & 3.44 & 1.53 \\
6981 &      M 72    &  2.28  &  4.60  & 2.38 & 1.21 \\
7089 &       M 2    &  1.08  &  3.56  & 4.00 & 1.59 \\
\hline
\end{tabular}
\end{table}

\subsection{Measuring the Stellar Mass Function} \label{criteria}

First, we used the available photometry to obtain estimates for 
the masses of the stars in our sample.  To do this, 
we fit theoretical isochrones taken from \citet{dotter07}
to the CMDs of every cluster.  Each isochrone
was generated using the metallicity and age of the cluster, and fit to
its CMD using the corresponding distance modulus and extinction
provided in \citet{dotter10}.
The MSTO was then defined using our isochrone fits by selecting the
bluest point along the main-sequence (MS).

We considered five stellar mass bins along the MS.  These ranged
from 0.25 - 0.75 M$_{\odot}$ in increments of 0.1 M$_{\odot}$.  This
range was chosen to help ensure complete sampling in all bins since the
lowest MSTO mass in our sample corresponds to $\approx$ 0.75 M$_{\odot}$,
and the photometric errors remain small ($\lesssim$ 0.05 mag) within
the magnitude range for each stellar mass bin in every cluster.  
We obtained number counts for
all stellar mass bins in the annulus r$_c <$ r $<$ 2r$_c$, where r is
the distance from the cluster center.  
This reduced our sample size by five clusters since the spatial coverage
offered by the ACS field of view is incomplete in these cases.  

We obtained completeness corrections for each stellar mass bin
in the annulus immediately outside the core (r$_c <$ r $<$ 2r$_c$).
This was done using the results of artificial star tests taken from
\citet{anderson08}.
Number counts for each mass bin were then multiplied by their
corresponding completeness corrections.  We did not include core
number counts in our analysis
since our completeness corrections begin to exceed 50\% somewhere
inside the core for every cluster in our sample.  This is due to
crowding and the high central surface brightnesses at the centres of
our clusters.  We have entirely removed three clusters
from our original sample used in \citet{leigh11}, namely NGC 1851,
NGC 5139, and NGC 6652.  This is because their completeness corrections
exceeded 50\% in every mass bin in the
annulus immediately outside the core.  We also removed additional
clusters from our samples for the lowest three mass bins whenever their
completeness corrections exceeded 50\%.  These clusters typically had
the highest MSTO masses.  In total, this left us with
27, 27, 23, 20, and 15 clusters in each of the five mass bins, in order of
decreasing stellar mass.  The completeness-corrected number counts for each 
stellar mass bin have been provided in Table~\ref{table:counts}.


\begin{table}
\centering
\caption{Completeness-corrected number counts for all five stellar mass 
bins in the annulus r$_c$ $<$ r $<$ 2r$_c$
  \label{table:counts}}
\begin{tabular}{|c|c|c|c|c|c|}
\hline
Cluster     &  MS1  &  MS2  &  MS3  &  MS4  &  MS5  \\
ID          &  (0.65 - 0.75 M$_{\odot}$)  &  (0.55 - 0.65 M$_{\odot}$)  &  (0.45 - 0.55 M$_{\odot}$)  &  (0.35 - 0.45 M$_{\odot}$) &  (0.25 - 0.35 M$_{\odot}$)  \\ 
\hline
    104     & 15113  & 16056  &    --  &    --  &    -- \\
   1261     &  4652  &  4840  &  4747  &  4647  &  4441 \\
   2298     &  1022  &   865  &   828  &   673  &   595 \\
   4147     &   525  &   355  &   257  &   154  &   100 \\
   4590     &  2830  &  3179  &  3793  &  4477  &  5543 \\
   5024     &  9773  &  9626  &  9725  &  8114  &  5859 \\
   5272     &  9763  & 10447  & 12097  & 12886  & 14339 \\
   5286     &  9394  &  8330  &  6436  &  2765  &   941 \\
   5904     &  5382  &  6726  &  9226  &    --  &    -- \\
   5927     &  4304  &  4208  &  5244  &  6043  &    -- \\
   5986     & 10328  & 10936  & 12070  & 13012  & 13635 \\
   6093     &  4356  &  2658  &    --  &    --  &    -- \\
   6121     &  1111  &   879  &    --  &    --  &    -- \\
   6171     &  1207  &  1049  &   968  &  1064  &    -- \\
   6205     & 11757  & 13012  & 16176  &    --  &    -- \\
   6218     &  2480  &  2337  &  2348  &  2589  &    -- \\
   6254     &  4631  &  4826  &  5375  &  6394  &  7342 \\
   6304     &  1806  &  1941  &    --  &    --  &    -- \\
   6341     &  4127  &  4019  &  3771  &  2456  &    -- \\
   6535     &   292  &   200  &   188  &   143  &   134 \\
   6584     &  3083  &  3330  &  3624  &  3880  &  4498 \\
   6637     &  3166  &  3192  &  3710  &  3818  &  1714 \\
   6779     &  3273  &  2941  &  2983  &  3047  &  3088 \\
   6838     &   592  &   634  &   654  &    --  &    -- \\
   6934     &  2615  &  2416  &  2228  &  1836  &  1302 \\
   6981     &  2731  &  2774  &  2914  &  2865  &  2847 \\
   7089     & 12549  & 12388  & 10113  &  3458  &    -- \\
\hline
\end{tabular}
\end{table}

\clearpage


The field of view of the ACS images is about
200'' on a side, which gives physical scales ranging between 1.5 and
16 pc (for the closest and furthest clusters in our sample).  Based on
this, we expect foreground contamination by field stars to be
negligible for most of the clusters in our sample given their current
locations in the Galaxy.  For example, \citet{dacosta82} considered
star count data 
in a similar area and over a comparable range of stellar masses for
three nearby globular clusters.  The author found that the 
corrections resulting from field contamination were always less than
10\% over nearly the entire range of stellar masses we are
considering.   

\subsection{Weighted Lines of Best-Fit} \label{lines}

In order to quantify cluster-to-cluster differences in the present-day 
stellar mass functions of the clusters in our sample, we obtained
lines of best-fit for 
(the logarithm of) the number of stars belonging to each stellar mass bin 
versus (the logarithm of) the total number of stars spanning all five
mass bins, which provides a proxy for the total cluster mass.  
This can be written:
\begin{equation}
\label{eqn:frac-bin}
\log N_{bin,i} = {\gamma_i}\log \Big( \frac{N_{tot}}{10^3} \Big) + \delta_i,
\end{equation}
where N$_{bin,i}$ is the number of stars belonging to mass bin $i$, N$_{tot}$ 
is the total number of stars spanning all five mass bins, and $\gamma_i$ 
and $\delta_i$ are both constants.  

Our motivation for adopting this technique is as follows.  If the 
fraction of stars belonging 
to each mass bin, or f$_{bin,i}$ $=$ N$_{bin,i}$/N$_{tot}$, 
is constant for all cluster masses, then we would expect N$_{bin,i}$ to 
scale linearly with N$_{tot}$.  Or, equivalently, $\gamma_i$ $\approx$ 1 in 
Equation~\ref{eqn:frac-bin}.  
However, if there is any systematic dependence of f$_{bin,i}$ on the
total cluster mass, then we should find that N$_{bin,i}$ does
\textit{not} scale linearly with N$_{tot}$.  In log-log space, the
slope of the line of best-fit for stellar mass bin $i$ 
should be less than unity (i.e. $\gamma_i$ $< 1$) if f$_{bin,i}$ 
systematically decreases with increasing cluster mass.  Conversely, 
we expect $\gamma_i$ $> 1$ if f$_{bin,i}$ systematically increases 
with increasing cluster mass.  This means that, for a sample of clusters 
with a wide range of total masses, we expect $\gamma_i < 1$ for 
the highest mass stars and $\gamma_i > 1$ for the lowest mass stars.  
This is because clusters lose preferentially low-mass stars due to 
two-body relaxation, and this process operates the fastest in lower 
mass clusters.

Equation~\ref{eqn:frac-bin} quantifies the number of stars belonging
to each stellar mass bin as a function of the total cluster mass.  More 
generally, it provides a means of quantifying cluster-to-cluster 
differences in the stellar mass function as a function of both the
stellar mass and the total cluster mass.  

The lines of best-fit have been weighted by adopting uncertainties for the
number of stars in each mass bin using Poisson statistics.  
Uncertainties for the slopes (i.e. for $\gamma_i$ in Equation~\ref{eqn:frac-bin}) 
were found using a bootstrap methodology in which we generated 1,000 fake
data sets by randomly sampling (with replacement) number counts from
the observations.  We obtained lines of best fit for each fake data
set, fit a Gaussian to the subsequent distribution and extracted its
standard deviation.  

\subsection{Monte Carlo Models} \label{models}

We have generated 268 Monte Carlo simulations for
globular cluster evolution spanning a range of initial total
numbers of stars, concentrations, virial radii and IMFs.  
The models realistically take into account 
both stellar and binary evolution, and track both short- and
long-range gravitational interactions between both single and binary
stars.  Detailed explanations concerning the development of these models 
can be found in \citet{joshi00}, \citet{joshi01}, \citet{fregeau03}, 
\citet{fregeau07}, and \citet{chatterjee10}.

For every combination of initial concentration (W$_0$), virial radius 
(r$_{vir}$) and IMF slope ($\alpha$ in Equation~\ref{eqn:kroupa}), we 
generated a series of models with different initial total numbers of 
stars (i.e. total cluster masses).  We adopted an IMF of the form:
\begin{equation}
\label{eqn:kroupa}
\frac{dN}{dm} = {\beta}m^{-\alpha},
\end{equation}
where $\alpha$ and $\beta$ are constants.  This was taken from 
\citet{kroupa01}, who fit a three-part 
power-law to this function with $\alpha = 2.3$ for 0.50 $<$
$m$/M$_{\odot}$ $<$ 1.00, $\alpha = 1.3$ for 0.08 $<$
$m$/M$_{\odot}$ $<$ 0.50, and $\alpha = 0.3$ for 0.01 $<$
$m$/M$_{\odot}$ $<$ 0.08.  We varied $\alpha$ only in the stellar 
mass range 0.08 $<$ $m$/M$_{\odot}$ $<$ 0.50.

Each model run was evolved for a period of 12 Gyrs, 
which roughly coincides with the ages of the clusters in our sample 
\citep[e.g.][]{deangeli05, marin-franch09}.  
The resulting collection of simulations spanned roughly the same range of 
total masses as our observed sample.  The initial cluster 
parameters considered in this paper are shown in 
Table~\ref{table:initial-conditions}.  With this suite of simulations, we 
have only scratched
the surface in terms of exploring the total initial parameter space that
could be relevant to the GCs in our observed sample.  However, our goal in this
paper is to demonstrate the strength of our technique for quantifying
cluster-to-cluster differences in the observed present-day MFs, and to show
by example how our method can be used to compare a large number of
observed MFs to analogous samples of simulated MFs.  We defer a more
complete exploration of the total possible 
parameter space of initial conditions to a future paper.

\begin{table}
\caption{Initial Model Parameters 
  \label{table:initial-conditions}}
\begin{tabular}{|l|c|}
\hline
Parameter                  &  Initial Values                  \\
\hline
IMF slope ($\alpha$)       &  1.3, 0.4, 0.0, -0.4                \\
Number of Stars            &  1e5, 2e5, 4e5, 6e5, 8e5, 1e6, 2e6, 4e6 \\
Concentration (W$_0$)      &  5.0, 5.5, 6.0, 6.5, 7.0             \\
Virial Radius (in pc)      &      3, 4, 5                       \\
\hline
\end{tabular}
\end{table}

The simulated clusters were placed on circular orbits at a distance 
of 4 kpc from the 
Galactic centre, and the resulting tidal effects from the Galaxy were 
accounted for.  We note that these effects are reduced by adopting 
a smaller initial virial radius.  The effects of tides 
were typically small in all but those models 
for which both the initial mass and concentration were very low, which 
agrees with the results of previous studies \citep[e.g.][]{vesperini97}.  
We assumed a metallicity of Z $= 0.001$ for all simulated clusters.  
This roughly agrees with what is typically observed in 
Galactic GCs \citep[e.g.][]{harris96}, and primarily
affects the rate of stellar mass loss due to winds early on in the
cluster lifetime when massive stars are still present 
\citep[e.g.][]{chernoff90}.  Although this does affect the rate of
dynamical evolution, the effect should be very similar from
cluster-to-cluster for the initial conditions considered in this paper.  
This is because the mass loss that occurs early on 
in the cluster lifetime due to stellar evolution has the greatest impact 
on clusters with very low initial concentrations, and can significantly 
reduce the time for cluster dissolution in these cases.  However, 
\citet{baumgardt03} showed that this effect is not severe for 
the range of initial concentrations (W$_0$ = 5 - 7) considered here.  
Finally, we assumed an initial global binary fraction of 10\% for all 
model runs using the same binary orbital parameter distributions as 
adopted in \citet{chatterjee10}.  We will return to these assumptions 
in Section~\ref{discussion}.

We generated simulated CMDs for every model run by converting the bolometric 
luminosity of every star to its corresponding magnitude in
the ACS F814W band.  This was done using the colour conversion routine of
\citet{pols98}, which uses the spectral libraries of \citet{lejeune97}
and \citet{lejeune98}.  For binary stars, the magnitudes of the components 
were combined in order to position them in the CMD as single objects.

Observations of star clusters are projected onto the plane of the sky,
whereas the output from our models provides only a 3-D distance from
the cluster centre for
every single and binary star.  Therefore, it was necessary to
convert these 3-D distances to corresponding 2-D values.  This was
done by randomly varying the component along the line-of-sight
to the cluster, and using the 3-D distance to calculate a
2-D value.  Using these projected 2-D distances from the cluster centre,
we also generated surface brightness profiles for every model run and
re-calculated a 2-D core radius (defined as the distance from the
cluster centre at which the surface brightness falls to half
its central value).  These 2-D core radii were then used to count 
the number of objects (i.e. single and binary stars) belonging to each stellar 
mass bin located within the annulus immediately outside the core 
(i.e. r$_c <$ r $<$ 2r$_c$).

\subsection{Comparing the Observed and Simulated Present-Day Mass Functions}

For every combination of initial concentration, virial radius, and IMF
slope, the different runs corresponding to different initial numbers of
stars were grouped together.  This gave us a sample of MFs spanning a range of 
total cluster masses for every combination of initial conditions.  The 
selection criteria described in 
Section~\ref{criteria} was then applied to each model, and lines of
best-fit were found
for each stellar mass.  The slopes of these lines of best-fit (i.e. $\gamma_i$ 
in Equation~\ref{eqn:frac-bin}) 
were then compared to the corresponding observed slopes for 
every stellar mass, and both the 
chi-squared value and the probability that the two samples (i.e. the observed
$\gamma_i$'s for all five mass bins and a given set of theoretically-derived 
$\gamma_i$'s for all mass bins) are drawn from the same distribution were found.  

\section{Results} \label{results}

In this section, we present the results of both our observational
analysis and its comparison to the models.

\subsection{Observational Analysis} \label{obs_analysis}

\begin{figure}
\begin{center}
\includegraphics[width=\columnwidth]{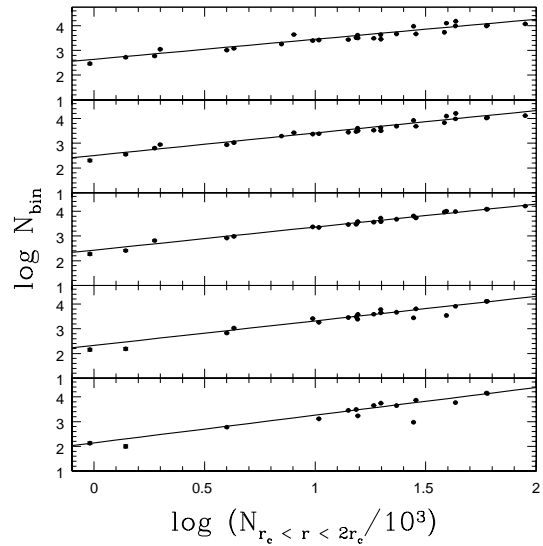}
\end{center}
\caption[Logarithm of the number of
stars belonging to each stellar mass bin as a function of the
logarithm of the total number of stars spanning all five mass
bins in the annulus immediately outside the core]{The logarithm of the number of
stars belonging to each stellar mass bin (N$_{bin}$) as a function of the
logarithm of the total number of stars spanning all five mass
bins in the annulus immediately outside the core (N$_{r_c < r < 2r_c}$).  In
descending order from top to bottom, the plots correspond 
to number counts in the mass ranges 0.65 - 0.75 M$_{\odot}$ (MS1),
0.55 - 0.65 M$_{\odot}$ (MS2), 0.45 - 0.55 M$_{\odot}$ (MS3), 0.35 -
0.45 M$_{\odot}$ (MS4), and 0.25 - 0.35 M$_{\odot}$ (MS5).  Lines of
best fit are shown for each mass bin by solid lines.
\label{fig:Ncore_vs_Nms_2rc}}
\end{figure}

\begin{figure}
\begin{center}
\includegraphics[width=\columnwidth]{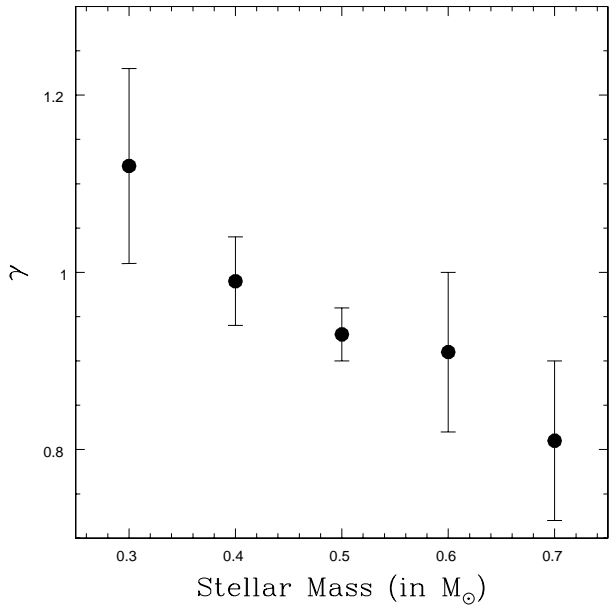}
\end{center}
\caption[Slopes for the lines of best-fit (i.e. $\gamma_i$) plotted as a 
function of stellar mass]{Slopes for the lines of best-fit, 
given by $\gamma_i$ in Equation~\ref{eqn:frac-bin}, plotted as a 
function of stellar mass.  
\label{fig:slopes}}
\end{figure}

We have plotted the logarithm of the number of stars in each stellar mass bin
versus the logarithm of the total number of stars spanning all five 
mass bins in Figure~\ref{fig:Ncore_vs_Nms_2rc}.  
The slopes and y-intercepts for the weighted lines of best-fit performed for 
each of these relations 
provided values for $\gamma_i$ and $\delta_i$ in Equation~\ref{eqn:frac-bin}.  
These are shown in Table~\ref{table:bestfit},
along with their corresponding uncertainties ($\Delta$$\gamma$ and 
$\Delta$$\delta$).  Each table entry has been 
provided in the form ($\gamma$ $\pm$ $\Delta$$\gamma$; $\delta$ $\pm$ 
$\Delta$$\delta$).  The values for $\gamma_i$ have also been plotted 
in Figure~\ref{fig:slopes}.  

\begin{table}
\centering
\caption{Lines of Best Fit for log N$_{bin,i}$ = ($\gamma$ $\pm$ $\Delta$$\gamma$)log (N$_{tot}$/10$^3$) + ($\delta$ $\pm$ $\Delta$$\delta$)
  \label{table:bestfit}}
\begin{tabular}{|l|c|}
\hline
         Stellar Mass            &          $\gamma$ $\pm$ $\Delta$$\gamma$; $\delta$ $\pm$ $\Delta$$\delta$         \\
\hline
MS1 (0.65-0.75 M$_{\odot}$)  &   0.81 $\pm$ 0.09; 2.64 $\pm$ 0.12   \\
MS2 (0.55-0.65 M$_{\odot}$)  &   0.91 $\pm$ 0.09; 2.50 $\pm$ 0.10   \\
MS3 (0.45-0.55 M$_{\odot}$)  &   0.93 $\pm$ 0.03; 2.43 $\pm$ 0.04   \\
MS4 (0.35-0.45 M$_{\odot}$)  &   0.99 $\pm$ 0.05; 2.33 $\pm$ 0.08   \\
MS5 (0.25-0.35 M$_{\odot}$)  &   1.12 $\pm$ 0.11; 2.14 $\pm$ 0.17   \\
\hline
\end{tabular}
\end{table}

As shown in Figure~\ref{fig:slopes}, $\gamma_i$ tends to
systematically increase with decreasing stellar mass.  The 
uncertainty for $\gamma_i$ is the highest 
for the lowest mass bin (MS5 in Table~\ref{table:bestfit}).  
This is because the photometric errors are the
highest at these dim magnitudes.  However, the errors are consistently 
at most $\approx$
10\% of the width in magnitude of their corresponding mass bin.  

In an attempt to improve
upon these statistics, we have also calculated reduced chi-squared
values with added intrinsic dispersion for the relations for each mass
bin.  That is, for each mass bin we added a constant term to the uncertainty for each
data point, found the uncertainty that yielded a reduced chi-squared of one,
and looked at the subsequent effects on the uncertainties for the
line of best-fit.  Based on this, we appear to be slightly over-estimating the
uncertainties for the MS1, MS2 and MS3 mass bins using our bootstrap
approach, and slightly under-estimating them for the MS4 and MS5
bins.  

\begin{figure}
\begin{center}
\includegraphics[width=\columnwidth]{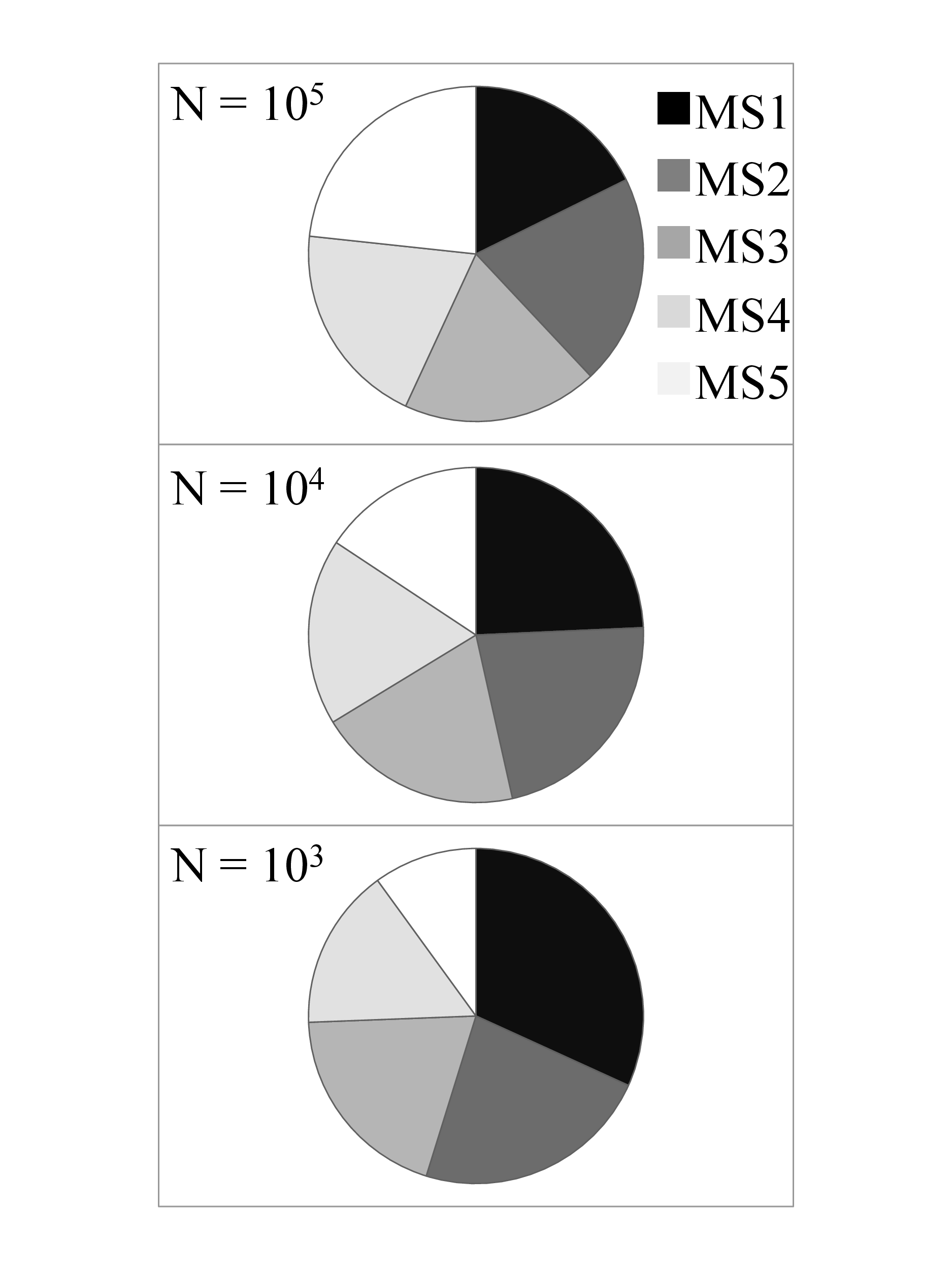}
\end{center}
\caption[Mass Functions in Pie Chart Form]{
  Stellar mass functions depicted in pie chart form.  The total
  area of each circle corresponds to the total number of stars
  spanning all five stellar mass bins, and each pie slice shows the
  fraction of this total corresponding to each mass bin.  Each of
  these fractions was calculated using the weighted lines of
  best-fit provided in Table~\ref{table:bestfit}.  From top to
  bottom, the total number of stars used to generate each pie chart
  was 10$^5$, 10$^4$, and 10$^3$.  Darker pie slices correspond to
  more massive bins.  The sequence of pie charts progressing from top
  to bottom effectively shows the evolution of 
  the stellar mass function with increasing dynamical age.
  \label{fig:pie_charts}}
\end{figure}

The change in the distribution of stellar masses as a function of the
total cluster mass can be illustrated using pie charts, as shown in
Figure~\ref{fig:pie_charts}.  Using the values for $\gamma_i$ and $\delta_i$ 
provided in Table~\ref{table:bestfit}, 
we have generated pie charts for three total numbers of stars
(spanning all five mass bins), namely N$_{tot}$ = 10$^5$, 10$^4$,
10$^3$ (from top to bottom in Figure~\ref{fig:pie_charts}).  As is
clear, low-mass stars become more and more preferentially depleted
with decreasing total cluster mass.  

From top to
bottom, the pie charts can be interpreted as depicting the evolution of 
the stellar mass function with increasing dynamical age.  This can be 
understood as follows.  
The inverse of the half-mass relaxation time can be used as a 
proxy for the rate of two-body relaxation throughout the entire 
cluster.  The half-mass relaxation time ranges from several 
million years to the age of the 
Universe or longer, and is approximated by \citep{spitzer87}:
\begin{equation}
\label{eqn:t-rh}
t_{rh} = 1.7 \times 10^5[r_h(pc)]^{3/2}N^{1/2}[m/M_{\odot}]^{-1/2}
years,
\end{equation}
where $r_h$ is the half-mass radius (i.e. the radius enclosing half
the mass of the cluster), $N$ is the total number of stars
within $r_h$ and $m$ is the average stellar mass.  
Simulations have shown that $r_h$ changes by a factor of at
most a few
over the course of a cluster's lifetime \citep{henon73, murray09}.
The GCs that comprise the ACS sample show a range of masses spanning
roughly 3 orders of magnitude (10$^4$-10$^6$ M$_{\odot}$), and have
comparably old ages ($\approx$ 10-12 Gyrs) \citep{deangeli05,
marin-franch09}.  Moreover,
their half-mass radii typically differ by less than a factor of 2 (see
Table~\ref{table:list}).  
Therefore, Equation~\ref{eqn:t-rh} suggests that the total cluster
mass provides a rough proxy for the degree of
dynamical evolution due to two-body relaxation.  In other words, the
effects of two-body relaxation on the evolution of
the stellar mass function should be the most pronounced in the least
massive clusters in the ACS sample \citep[e.g.][]{demarchi07,
  baumgardt08, kruijssen09}.  Said another
way, dynamical age increases with decreasing cluster mass.  Therefore, 
Figure~\ref{fig:pie_charts} shows that our results are consistent with 
the general picture that two-body relaxation 
is the cause of the observed depletion of low-mass stars in low-mass
clusters, as opposed to some unknown feature of the star formation
process. 

\subsection{Theoretical Analysis} \label{theory_analysis}

In this section, we compare the results of our observational analysis
to 268 Monte Carlo simulations for globular
cluster evolution spanning a range of initial conditions.  

\begin{figure}
\begin{center}
\includegraphics[width=\columnwidth]{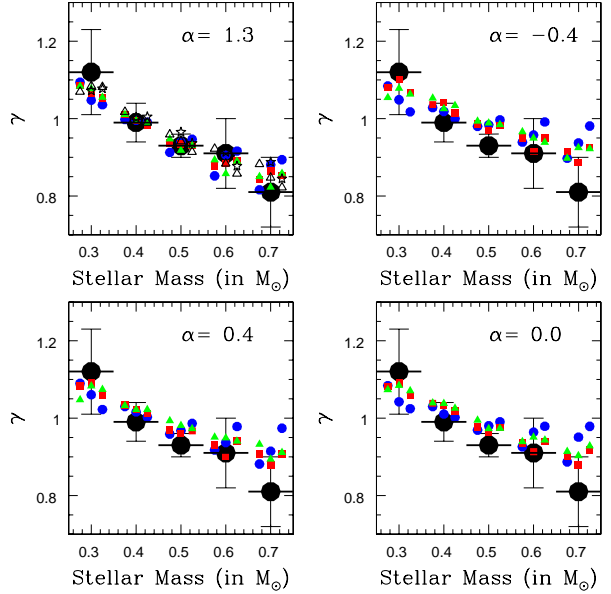}
\end{center}
\caption[Comparison of the observed $\gamma_i$ values to 
the corresponding model $\gamma_i$ values for every combination of initial 
conditions]{Comparison of the observed $\gamma_i$ values 
to the corresponding model $\gamma_i$ values for every combination of initial 
conditions.  The observed $\gamma_i$'s are shown as solid circles, whereas 
the model $\gamma_i$'s are shown as different symbols.  Each inset shows the model 
$\gamma_i$'s for a different IMF.  Starting at the upper right 
and rotating clockwise, the insets correspond to IMF slopes 
(in Equation~\ref{eqn:kroupa}) of -0.4, 0.0, 0.4, and 1.3.  The online 
version of this plot shows the model $\gamma_i$'s either as open or coloured 
symbols which 
have been used to indicate different initial concentrations.  The solid blue 
circles, solid red squares, solid green triangles, open black triangles, and 
open five-point stars correspond to initial concentrations (W$_0$) of 5.0, 5.5, 
6.0, 6.5, and 7.0, respectively.  A small offset has been implemented for 
every stellar mass bin to indicate different initial virial radii, with 
the virial radius increasing from left to right.  A horizontal line has also 
been included on every observed data point in order to indicate the 
width of the corresponding stellar mass bin.
\label{fig:slopes-compare}}
\end{figure}

Figure~\ref{fig:slopes-compare} shows a comparison between all five 
$\gamma_i$ values found from the observed MFs and the corresponding 
model $\gamma_i$ values for every combination of initial conditions.  
As is clear, the agreement is excellent for nearly every combination of 
initial IMF, concentration and virial radius.  This was confirmed by 
our chi-squared values, and the probability that the observed and 
model $\gamma_i$'s are drawn from the same distribution exceeded 64\% 
for all comparisons.  

Our results suggest that a Kroupa IMF (i.e. $\alpha = 1.3$ in 
Equation~\ref{eqn:kroupa}) typically gives the best agreement with the 
observations.  Every set of models with this IMF yielded a probability 
greater than 93\% that the observed and model 
$\gamma_i$'s are drawn from the same distribution.  
Our results also appear to be relatively 
insensitive to the initial concentration and virial radius.  This agrees 
with what was found by \citet{vesperini97} and \citet{baumgardt03} given 
the limited ranges we have explored for these parameters.

\section{Summary \& Discussion} \label{discussion}

In this paper, we have presented a new technique to quantify
cluster-to-cluster variations in a large sample of observed stellar mass
functions.  Our method quantifies these differences as a function of both
the stellar mass and the total cluster mass, and offers the advantage 
that it is insensitive to 
the exact functional form of the MF.  We have applied our technique to
completeness-corrected stellar mass 
functions in the range 0.25-0.75 M$_{\odot}$ for a sample of 27
globular clusters taken from the ACS Survey for GCs, 
and have compared the results to a series of Monte Carlo models for 
GC evolution.

In the subsequent sections, we discuss the implications of our results for the formation 
and evolution of Milky Way GCs.

\subsection{The Effects of Two-Body Relaxation} \label{2-body}

We have shown that the observed
differences in the present-day MFs in our sample can be reproduced
by assuming (1) an universal initial mass function for all clusters, and 
(2) that internal two-body relaxation is the dominant mechanism 
contributing to the cluster-to-cluster variations.  

Our results are 
the most reliable in the mass range 0.45 - 0.75 M$_{\odot}$.  This 
is due to the larger photometric errors at the fainter magnitudes 
corresponding 
to lower stellar masses, and incompleteness resulting from crowding.  
Despite the high quality of
the data used in this study, these issues are currently unavoidable
given the nature of the observations.  This will be a key challenge
for future studies to resolve, however the method we have presented
in this paper offers a robust means of performing future analyses.  
We note that the results of our observational analysis
are consistent with those of \citet{demarchi10}, who fit a tapered
power-law distribution function with an exponential truncation to the
stellar mass functions of a sample of 30 clusters containing both
young and old members.  We have also verified that our results are 
consistent with those of 
\citet{paust10}, who performed power-law fits to the MFs of 17 GCs 
taken from the ACS Survey.

In this paper, we have focused on the \textit{local}
MFs in the central cluster regions of the GCs in our sample.  However,
previous studies have shown that the \textit{local} MF can differ
considerably from the \textit{global} MF \citep[e.g.][]{vesperini97}.
In principle, this should not have affected our comparisons
between the observed and simulated MFs since we have considered the same
structural areas of the clusters in all cases.

\subsection{The Effects of Binary Stars} \label{binaries}

Binary stars are unresolved in GCs, appearing as single
objects located above the MS in the
cluster CMD.  Therefore, we have included some objects in our number 
counts that are in fact binaries masquerading as single stars.  
Previous studies have shown that unresolved binaries can contribute to
flattening, or even inverting, the stellar mass function in the range
of stellar masses considered in this paper \citep[e.g.][]{marks08, marks10}.  
Moreover, observational evidence suggests that 
the binary fractions in GCs are inversely proportional to the total
cluster mass \citep[e.g.][]{sollima08, milone08, knigge09}.  
In particular, the most massive MW GCs tend to have binary fractions on the order of
only a few percent \citep[e.g.][]{rubenstein97, cool02, davis08}, whereas
the least massive MW GCs tend to have larger binary fractions that can
even exceed 50\% in some cases \citep[e.g.][]{milone11}.  
This suggests that unresolved binaries should have had the 
largest effect on the MFs of the lowest mass clusters in our sample.  
Therefore, unresolved binaries could also be contributing to the general 
trend we have found of increasing $\gamma_i$ with 
decreasing stellar mass.  

In an effort to quantify the effects of unresolved binaries on our results, 
we removed all binaries from our simulated MFs and re-performed our 
weighted lines of best-fit.  This confirmed that unresolved binaries 
could indeed be contributing to the general trend we have found of 
increasing $\gamma_i$ with decreasing stellar mass.  
However, this effect was not significant in the models, in large part 
because we assumed a constant initial global binary fraction of 10\% 
for all clusters.  
In order to reproduce the observed trend in $\gamma_i$, it would require 
that the range in binary fractions between low- and high-mass clusters 
is greater than the observed range by a factor $\gtrsim$ 2 
\citep[e.g.][]{milone11}.  Although the binary fractions generally 
increase in the core over time in our simulations (see \citet{fregeau09} 
for more details), they are not sufficiently high to have 
significantly contributed to the trend of increasing $\gamma_i$ with 
decreasing stellar mass.  At the end of our simulations, the core 
binary fractions are typically in the range 10-30\%.

In order to properly assess these effects,
objects that are in fact unresolved binaries should be identified and our
analysis of the observations should be re-performed.  This could
be done using multi-band photometry since, if a given binary happens to
fall on a single star evolution track in one CMD, it is unlikely to fall
on the corresponding tracks in
other CMDs constructed using different wavelength bands.  Stellar
evolution models could then be used to constrain the masses of the
component stars.  We intend to address this issue in future work.

We expect that the influence of binaries on GC evolution
by, for example acting as heat-sources via hardening encounters
with single stars \citep[e.g.][]{hut83a, hut83b, fregeau09}, should
have had a negligible
impact on our results.  This is because most of the clusters
in our sample should still be undergoing core contraction
\citep[e.g.][]{gieles11}.  It follows that their central densities
have not yet
become sufficiently high for encounters involving binaries to occur
frequently enough that they could have significantly affected the cluster
evolution.  Notwithstanding, future studies should incorporate
models spanning a range of realistic initial binary fractions and distributions
of orbital parameters in order to properly assess all of these effects.

\subsection{The Effects of Tides from the Galaxy} \label{tides}

Tides from the Galaxy effectively reduce the time-scale
on which two-body relaxation operates \citep[e.g.][]{heggie03}.  
This primarily serves to make clusters appear more dynamically evolved 
than they otherwise would.  The same effect is caused by disc shocking 
which, as with tidal effects, should most severely affect clusters with 
the lowest masses and the smallest Galactocentric distances.  Therefore, 
the locations of clusters within their host galaxies has been shown to 
play an important role in determining the degree of flattening of their 
stellar MFs \citep[e.g.][]{vesperini97}.  In an effort to quantify the
effects caused by tides on our observational results, 
we performed several cuts in perigalacticon distance 
and re-performed our weighted lines of best-fit.  Estimates for the 
perigalacticon distances were obtained from \citet{dinescu99} and 
\citet{dinescu07} for every cluster in our sample.  Despite
removing clusters with small perigalacticon distances
for which it is typically argued that tidal effects should be the most
severe \citep[e.g.][]{heggie03}, our lines of best-fit
remained the same.  We caution that these effects 
have not been fully accounted for in the simulations of GC evolution 
performed in this paper, for which we adopted circular orbits and a 
Galactocentric distance of 4 kpc in all cases.  We intend to adopt 
a more realistic distribution of Galactic orbits for our model clusters 
in a future paper in order to properly assess the effects caused by tides 
in determining the present-day MFs of the GCs in our observed sample.

\subsection{The Effects of Primordial Gas Expulsion} \label{gas-expulsion}

As discussed in \citet{marks08} and \citet{marks10}, the expulsion of primordial 
gas early on in the cluster lifetime can have a dramatic effect on the stellar 
MFs of clusters.  In particular, these authors showed that 
clusters that began their lives with smaller concentrations are more likely 
to have lost a larger fraction of their low-mass stars as a result of this 
effect.  
Primordial gas expulsion could therefore contribute to improving the agreement we
have found between the observed and simulated MFs.  This would be
accomplished if our $\gamma_i$ values for the models were 
simultaneously increased for low stellar masses and decreased for high
stellar masses, as is evident from Figure~\ref{fig:slopes-compare}.  This
would occur if primordial gas expulsion had a more pronounced effect on
the MFs of the lowest mass clusters in our sample.  
This is not unreasonable since the depth of
the cluster potential increases with increasing cluster mass.

\subsection{Future Work} \label{future}

Given the very old ages and 
therefore low metallicities ([Fe/H] $\approx$ -2.28 -
(-0.37)) of the clusters in our sample, our technique could potentially 
be used to better constrain the IMFs of old massive 
star clusters and, more generally, star formation in the very early 
Universe.  This will be done in a forthcoming paper 
by considering a larger range of IMFs and initial cluster 
conditions than we have considered here.

Finally, we wish to point out that the method we have presented 
can be generalized to compare any large sample of distribution functions.  
We intend to illustrate this in a 
future paper by using our technique to quantify cluster-to-cluster
differences in the orbital distributions (period, eccentricity, and
mass-ratio) of the binary populations in GCs as a function of the
total cluster mass.

\section*{Acknowledgments}

We would like to thank an anonymous referee for several 
suggestions that helped to improve our manuscript.  We also 
wish to thank Christopher McKee for useful discussions, Aaron Dotter 
and Roger Cohen for their support in analyzing the observations, and 
Robert Cockcroft for a critical read of our manuscript.  N.L. was 
supported by Ontario Graduate Scholarships (OGS) and the European 
Space Agency (ESA).  S.U. was supported by NSF Grant AST-0607498 
and NASA ATP Grant NNX09AO36G at Northwestern University.


\bsp

\label{lastpage}

\end{document}